\documentclass{ifacconf}
\usepackage{graphicx}      
\usepackage{natbib}        
\usepackage{amsmath}
\usepackage{amsfonts}
\usepackage{subcaption}
\usepackage{graphicx}
\usepackage{multirow}
\usepackage{ amssymb }

\begin{document}
\begin{frontmatter}

\title{Initialization Approach for Nonlinear State-Space Identification via the Subspace Encoder Approach} 


\author[First]{Rishi Ramkannan}  
\author[First]{Gerben I. Beintema}  
\author[First,Second]{Roland T\'{o}th}  
\author[First]{Maarten Schoukens}  

\address[First]{Control System group, Eindhoven University of Technology, Eindhoven, the Netherlands}
\address[Second]{Systems and Control Laboratory, Institute for Computer Science and Control, Budapest, Hungary}

\begin{abstract} 
The SUBNET neural network architecture has been developed to identify nonlinear state-space models from input-output data. To achieve this, it combines the rolled-out nonlinear state-space equations and a state encoder function, both parameterised as neural networks The encoder function is introduced to reconstruct the current state from past input-output data. Hence, it enables the forward simulation of the rolled-out state-space model. While this approach has shown to provide high-accuracy and consistent model estimation, its convergence can be significantly improved by efficient initialization of the training process. This paper focuses on such an initialisation of the subspace encoder approach using the Best Linear Approximation (BLA). Using the BLA provided state-space matrices and its associated reconstructability map, both the state-transition part of the network and the encoder are initialized. The performance of the improved initialisation scheme is evaluated on a Wiener-Hammerstein simulation example and a benchmark dataset. The results show that for a weakly nonlinear system, the proposed initialisation based on the linear reconstructability map results in a faster convergence and a better model quality. 
\end{abstract}


\end{frontmatter}


\section{Introduction}
    Mathematical models are essential to understand the dynamical behaviours of engineering systems. These models are utilised for control design, fault diagnosis, or the prediction or simulation of systems. Linear system identification techniques \citep{ljung_sysid,pintelon2012system} have been successfully employed to obtain black-box linear models of systems starting from input-output data. However, technological advances to meet industrial and consumer demands are driving system designs towards nonlinear operational regimes. However, the nonlinear system identification field is less mature compared to its LTI counterpart \citep{Schoukens2019overview}. 
    
    While there is a wide range of nonlinear model classes, this paper focuses on nonlinear state-space (NLSS) identification \citep{Suykens1995,paduart2010identification,schoukens2021improved}. NLSS identification requires a potentially non-convex nonlinear optimization problem to be solved. A good initialisation of the parameter estimates could lead to faster convergence of the optimization algorithm and a higher likelihood of converging to the global minimum. 
    
    The SUBNET artificial neural network (ANN) architecture proposed in \citep{beintema2021nonlinear} and studied in detail in \citep{beintema2022} has proven to offer a versatile and robust (nonlinear) system identification approach over a wide range of model classes and applications from nonlinear state-space, to Koopman and linear parameter-varying identification \citep{beintema2021nonlinear,beintema2021non,Iacob2021,Verhoek2022}. It combines an improved computational efficiency, increased cost smoothness and utilizes effective nonlinear optimization approaches. Nevertheless, as the parameters of the SUBNET network in the corresponding estimation scheme are currently initialized randomly, a reliable parameter initialization could further improve the model quality and/or time required for the optimization approach to converge. This is illustrated in a wide range of earlier approaches, one of the most common approaches of initialisation of black-box nonlinear models is by using a linear approximation of the system. This approach has proven to be effective over a wide range of model structures including block oriented nonlinear models \citep{schoukens2017identification}, linear fractional representation-based nonlinear models \citep{schoukens2020initialization}, Polynomial NLSS models \citep{paduart2010identification} and state space models parameterised as ANNs \citep{Suykens1995,schoukens2021improved}.
    
    This paper investigates the initialisation of the parameters present in the SUBNET architecture when used for nonlinear state-space neural identification. Three initialization schemes are compared. The state and output equations are initialized randomly or based on the BLA state-space matrices similar to \citep{schoukens2021improved}. The encoder network present in the SUBNET architecture is either randomly initialized or initialized using the reconstructability map obtained from the BLA model of the system under test. The performance of the proposed initialisation approach is analysed on a Wiener-Hammerstein simulation system and the well-established Wiener-Hammerstein benchmark system \citep{schoukens2009wiener}. The results show that for a weakly nonlinear system, the proposed initialisation scheme results in a faster convergence and a better model quality. 
    
    The remainder of the paper starts with the introduction of the considered system and the model class in Section~\ref{ses:System and model}. An overview of the subspace encoder method is given in Section~\ref{Ses:Subspace}. Section~\ref{ses:problem} describes the proposed initialisation based upon the BLA estimate. The proposed initialisation schemes are tested using a simulation study and the conclusions are drawn in Sections \ref{ses:results} and \ref{sec:conclusion} respectively.

\section{System and Model class} \label{ses:System and model}
The fading memory nonlinear discrete-time systems class that can be represented in the state-space form is considered:
\begin{subequations}\label{equ:system}
\begin{align}
    x_{t+1} &= f(x_t,u_t)\label{equ:systemf}  \\
    y_t &= h(x_t,u_t) + v_t\label{equ:systemh},
\end{align}
\end{subequations}
where \eqref{equ:systemf} is the nonlinear state equation and \eqref{equ:systemh} is the nonlinear output equation, $u_t \in \mathbb{R}^{n_u}$ is the system input, $y_t \in \mathbb{R}^{n_y}$ is the noisy system output, $x_t \in \mathbb{R}^{n_x}$ is the internal states, and $v_t  \in \mathbb{R}^{y}$ represents an external, possibly colored, additive noise source with finite variance. 

The objective of this paper is to estimate a nonlinear discrete-time state space model of~\eqref{equ:system} starting from data generated by~\eqref{equ:system}. The considered nonlinear state-space model structure for this task is described below,
\begin{subequations}\label{equ:model}
\begin{align}
    \hat{x}_{t+1} &= A \hat{x}_t + B u_t + f_{\theta_\text{NL}}(\hat{x}_t,u_t), \label{equ:model-state-equ} \\
    \hat{y}_t &= C \hat{x}_t + D u_t + h_{\theta_\text{NL}}(\hat{x}_t,u_t), \label{equ:model-output-equ}
\end{align}
\end{subequations}
where $A\in\mathbb{R}^{n_x\times n_x}$, $B\in\mathbb{R}^{n_x\times n_u}$, $C\in\mathbb{R}^{n_y\times n_x}$ and $D\in\mathbb{R}^{n_y\times n_u}$ are matrices describing the linear model terms. The functions $f_{\theta_\text{NL}}$ and $h_{\theta_\text{NL}}$ are static nonlinear functions parameterised as fully connected multi-layer ANNs. Furthermore, $\hat{x}_t$ is the modelled state, $\hat{y}_t$ is the model output and, $\theta = \mathrm{vec}(A, B, C, D, \theta_\text{NL})$ represents the model parameters collected in a vector. In \citep{schoukens2021improved}, it has been shown that having an explicit linear part in parallel to the nonlinear part for state-space ANN model structure results in improved training behaviour. However, note that introducing such an explicit linear part does not alter the class of systems represented by \eqref{equ:model}. Indeed, the linear parts can easily be absorbed in the nonlinear functions $f_{\theta_\text{NL}}$ and $h_{\theta_\text{NL}}$ which are parameterized by ANNs. In the remainder of the paper, for notational simplicity, the direct feed-through term from the input to the output is dropped. Finally, note that the considered model representation is not unique. There could be multiple values of $\theta$ for which the same input-output behaviour is obtained, this is a common issue in black-box nonlinear state-space identification. 

\section{Subspace Encoder-Based Identification} \label{Ses:Subspace}

The subspace encoder identification approach introduced in \citep{beintema2021nonlinear} combines the use of multiple shooting with an encoder function that estimates the initial state from past inputs and outputs. It is shown that multiple shooting smoothens the loss landscape, improving the parameter estimation \citep{ribeiro2020smoothness}. This method involves splitting the data set into multiple (short) sections and computing the loss independently over these sections. This results in the following loss (identification cost function) evaluated along the given data-set:
    \begin{subequations}
    \begin{align}\label{Equ:total_loss}    
        V(\theta) &= \frac{1}{M} \sum_{t=n+1}^{N-T+1} \sum_{k = 0}^{T-1} || \hat{y}_{t+k|t} - {y}_{t+k}||^{2}_{2}, \\ 
        \hat{y}_{t+k|t} &= C \hat{x}_{t+k|t} +  h_{\theta_\text{NL}}(\hat{x}_{t + k|t},u_{t+k}), \label{eq:LossOutputFunction}\\
        \hat{x}_{t+k+1|t} &=  A\hat{x}_{t+k|t} + B u_{t+k} + f_{\theta_\text{NL}}(\hat{x}_{t+k|t},u_{t+k}), \label{eq:LossStateFunction}\\
        \hat{x}_{t|t} &= W_u u_{t-n_b:t-1} + W_y y_{t-n_a:t-1} \label{eq:encoderFunction} \\ 
        & \quad + \psi_{\theta_\text{NL}}(y_{t-n_a:t-1},u_{t-n_b:t-1}), \nonumber
    \end{align}
   \end{subequations}
where~\eqref{eq:LossOutputFunction} and~\eqref{eq:LossStateFunction} provide the forward simulation of the model and~\eqref{eq:encoderFunction} determines, or encodes, the initial state from past input output data. Furthermore, $M = (N - T - n + 1) T$, $T$ denotes the number of steps in the future for which the simulation is performed given the initial time index $t$ and the pipe ($|$) notation is used to distinguish between different subsections as $(\text{current index} | \text{start index})$, while $u_{t-n_b:t-1} \triangleq [u_{t-n_b}^\top,...,u_{t-1}^\top]^\top$, $y_{t-n_a:t-1} \triangleq [y_{t-n_a}^\top,...,y_{t-1}^\top]^\top$. The loss for each subsection can be calculated in parallel, which allows for the use of mini-batching during optimization. Note that, even though at each optimization step only short subsections of the complete dataset are considered, during the optimization the full dataset is used as at each optimization step new random subsections are selected to compute the gradient.

\begin{figure}[htb]
\begin{center}
\includegraphics[width=8.4cm]{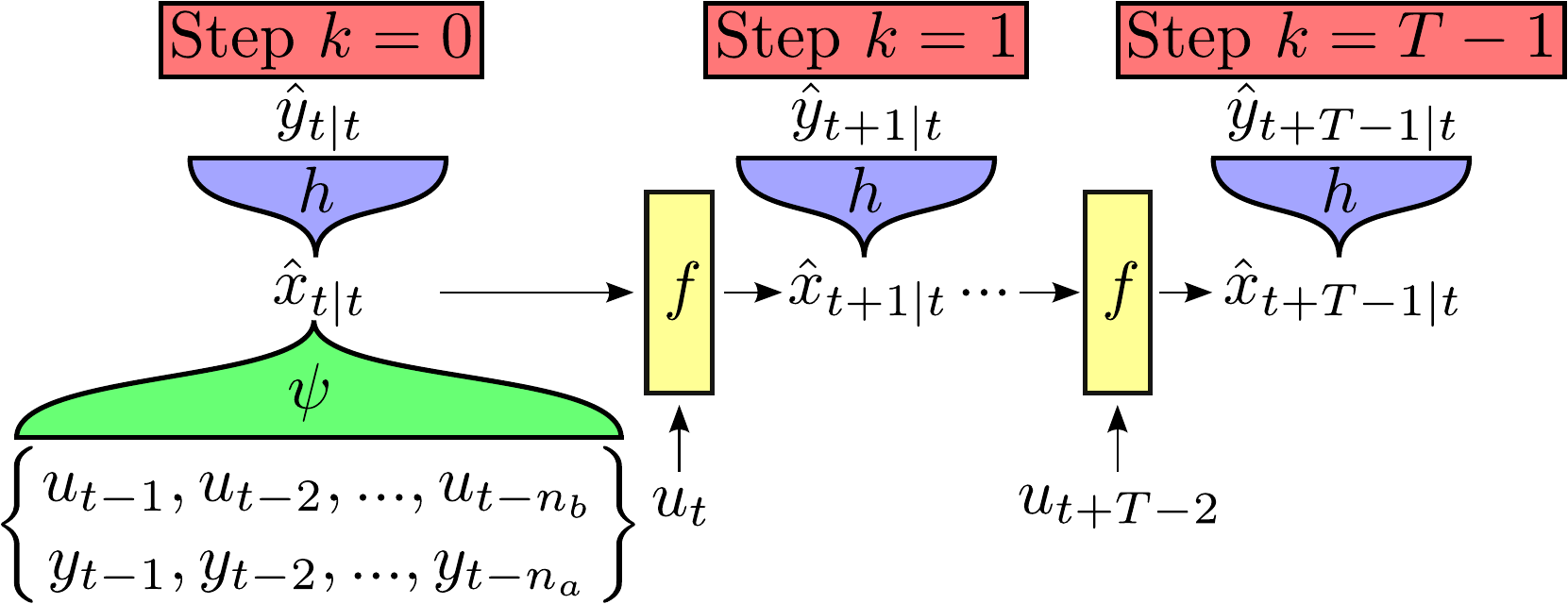}    
\caption{Structure of the subspace encoder network (SUBNET).} 
\label{fig:encoder}
\end{center}
\end{figure}

Estimating the initial states at the start of each section plays a crucial role in subspace encoder-based identification. The encoder function $\psi_{\theta_\text{NL}}$ given in \eqref{eq:encoderFunction}, parameterised as an artificial neural network, is utilised to obtain the initial states $\hat{x}_{t|t}$ during training. It acts as a reconstructability map, since, it obtains the initial state from the past inputs $u_{t-n_b:t-1}$, and outputs $y_{t-n_a:t-1}$ where $n_a$ and $n_b$ denote the maximum past time lags of the outputs and inputs used by the encoder. $W_u$, $W_y$ and $\psi_{\theta_{NL}}$ are jointly estimated with the state-transition and output functions using the loss~\eqref{Equ:total_loss}. The resulting unrolled state-space neural network combined with the encoder network is visualized in Figure~\ref{fig:encoder}.

We refer the reader to \citep{beintema2021nonlinear} for a more detailed description of the subspace encoder approach.

\section{Improved Subspace Encoder Initialization}\label{ses:problem}

The subspace encoder approach in \citep{beintema2021nonlinear} utilises random parameter initialization using the uniform distributions given by the Xavier initialization~\citep{glorot2010Xavierinit} for the state, output and encoder functions. Although, Xavier initialization is commonly used within ANN training, the nonlinear optimization problem remains prone to local minima or possibly long optimization times. Providing a better initial estimate, can reduce the required optimization time and/or improve the quality of the resulting models. The subsections below describe how the BLA and its associated reconstructability map can be utilized to provide an improved initialization for the subspace encoder approach.

\subsection{Best Linear Approximation}
The Best Linear Approximation (BLA) provides a linear time invariant (LTI) approximation of a nonlinear system. The BLA is best in a mean square sense for the class of chosen input signals \citep{pintelon2012system}:
\begin{subequations}
    \begin{equation}
        G_{BLA}(q) = \underset{G(q)}{\mbox{arg min}}\ E_{u,v}\{\|\Tilde{{y}}_t - G(q)\Tilde{{u}}_t\|^2_2\},
    \end{equation}
    \begin{align}
        \Tilde{{u}}_t &= u_t - E_{u}\{u_t\}, \\
        \Tilde{{y}}_t &= y_t - E_{u,v}\{y_t\},
    \end{align}
\end{subequations}   
where $E_{u,v}$ denotes the expectation operator taken w.r.t the random variations due to the input realizations of $u_t$ and the output noise $v_t$ and $G(q)$ belongs to the set of all possible discrete-time LTI systems. Practically, a BLA estimate can be obtained by classical prediction-error LTI state-space identification approaches \citep{ljung_sysid,pintelon2012system}. Without loss of generality, in the remainder of the paper we assume that the input and output signals are zero-mean and normalized during data preprocessing. 

The BLA can be estimated as a linear state-space model resulting in the state-space matrices ($\tilde{A}$, $\tilde{B}$, $\tilde{C}$):
    \begin{subequations}\label{equ:BLA-ss}
    \begin{align}
        \hat{x}^{BLA}_{t+1} &= \tilde{A}\hat{x}^{BLA}_{t} + \tilde{B}\tilde{u}_t, \label{equ:BLA-state}\\
        \hat{y}^{BLA}_t &= \tilde{C}\hat{x}^{BLA}_{t}. \label{equ:BLA-output}
    \end{align}
    \end{subequations}
These matrices will later be used to initialize the subspace encoder model estimate. It is recommended, if possible, to use the same data to obtain the BLA estimate as for the identification of the nonlinear state space model. This ensures that the linear approximation is valid in the data range of interest that is considered for the nonlinear identification. 

\subsection{Reconstructability Map} 
By time-inverting of the BLA linear SS equations Eq. \eqref{equ:BLA-ss}, with no direct feedthrough ($D = 0$), the past outputs described as (see \citep{callier2012linear});
\begin{equation}\label{eq:ypast-given-xt-ut}
    \hat{y}^{BLA}_{t-n:t-1} = [\tilde{C}\tilde{A}^-]_\text{map} \hat{x}^{BLA}_t - [\tilde{C}\tilde{A}^-\tilde{B}]_\text{map} \tilde{u}_{t-n:t-1},
\end{equation}
where $\tilde{A}^-$ is used to indicate that inverses of the state matrix are involved in constructing the map. The maps are given by 
\begin{subequations}
\begin{equation}\label{Equ:CA}
    [\tilde{C}\tilde{A}^-]_\text{map} = \begin{bmatrix}
        \tilde{C}\tilde{A}^{-n}\\
        \vdots\\
        \tilde{C}\tilde{A}^{-1}
        \end{bmatrix}, 
\end{equation}
\begin{equation}\label{Equ:CA CAB}
    [\tilde{C}\tilde{A}^-\tilde{B}]_\text{map} = \begin{bmatrix}
             \tilde{C}\tilde{A}^{-n}\tilde{B} & \tilde{C}\tilde{A}^{1-n}\tilde{B}  & \ldots &\tilde{C}\tilde{A}^{-1}\tilde{B} \\
            \vdots & \vdots& \ddots  & \vdots\\
            \tilde{C}\tilde{A}^{-2}\tilde{B} & \tilde{C}\tilde{A}^{-1}\tilde{B} & \ldots  & 0\\
            \tilde{C}\tilde{A}^{-1}\tilde{B} & 0 & \ldots & 0
            \end{bmatrix}.
\end{equation}
\end{subequations}
The reconstructability map is obtained by solving the possibly over-determined system of equations~\eqref{eq:ypast-given-xt-ut} for $\hat{x}_t$ using the left pseudo inverse of $[\tilde{C}\tilde{A}^-]_\text{map}$ given by $[\tilde{C}\tilde{A}^-]_\text{map}^\dagger$. Hence, the initial state of the simulation model can be recovered as  
\begin{equation} \label{equ:map}
    \hat{x}^{BLA}_t = [\tilde{C}\tilde{A}^-]_\text{map}^\dagger \left ( \hat{y}^{BLA}_{t-n:t-1} + [\tilde{C}\tilde{A}^-\tilde{B}]_\text{map} \tilde{u}_{t-n:t-1} \right ).
\end{equation}

In general, for systems of order $n_x$, at least $n \geq n_x$ past input-output samples are necessary for the existence of a unique pseudo inverse of $[\tilde{C}\tilde{A}^-]_\text{map}$. Thus this is also necessary for the uniqueness of the reconstructability map. Furthermore, observability of the obtained BLA model is also a necessary condition~\citep{callier2012linear}.

\subsection{Proposed BLA Parameter Initialisation} \label{Ses:init}

The initialisation of the state, output and encoder network using the BLA and the reconstructability map is performed by initialising the weights and biases under the assumption that the inputs to the network are normalised to have a zero mean and a unit standard deviation. To initialise the state-space and encoder function with the BLA, 
we rewrite the nonlinear functions $f_{\theta_\text{NL}}, h_{\theta_\text{NL}}, \psi_{\theta_\text{NL}}$ in the form
\begin{subequations}
\begin{align*}
    f_{\theta_\text{NL}}(\hat{x}_{t+k|t},u_{t+k}) &= W_\text{last}^f \phi_{\theta_\text{NL}}^f( [\hat{x}_{t+k|t}^\top u_{t+k}^\top]^\top) + b_\text{last}^f \\
    h_{\theta_\text{NL}}(\hat{x}_{t+k|t}) &= W_\text{last}^h \phi_{\theta_\text{NL}}^h( \hat{x}_{t+k|t} ) + b_\text{last}^h
\end{align*}
\begin{equation*}
    \psi_{\theta_\text{NL}}(y_{t-n_a:t-1},u_{t-n_b:t-1}) = W_\text{last}^\psi \phi_{\theta_\text{NL}}^\psi(.,.) + b_\text{last}^\psi
\end{equation*}
\end{subequations}
where $\phi$ indicates the output of the last hidden layer after activation. Writing these functions in this form allows us to ``turn off'' their influence by setting the parameters $W_\text{last}$ and $b_\text{last}$ to zero. Setting these terms to zero will ensure that the initial model behaves like the BLA estimate. All other ANN layer weights are randomly initialized.

Considering the subspace encoder approach, the initialisation scheme using the BLA of the system obtained based on the dataset can be utilised for the state and the output networks by setting  $A = \tilde{A},\ B = \tilde{B},\ C = \tilde{C}$, and $W_\text{last}^f = W_\text{last}^h =0$, $b_\text{last}^f = b_\text{last}^h =0$, similar to \citep{paduart2010identification,schoukens2020initialization,schoukens2021improved}.

However, the encoder network also plays a crucial role by estimating the initial state for each subsection. The encoder can be initialized using the BLA estimate by setting $W_u = [\tilde{C}\tilde{A}^-]_\text{map}^\dagger [\tilde{C}\tilde{A}^-\tilde{B}]_\text{map}$, $W_y = [\tilde{C}\tilde{A}^-\tilde{B}]_\text{map}^\dagger$ and $W_\text{last}^\psi = 0$, $b_\text{last}^\psi = 0$. Again, setting the last layer terms to zero ensures that the encoder behaves like the BLA reconstructability map after initialization. This ensures that the encoder approximately reconstructs the state of the BLA estimate. However, this will not be exact as the measured system output is used when evaluating the encoder instead of simulated BLA output as is done in~\eqref{equ:map}.

By combining the different initialization options three different initialization schemes are obtained: 1) a fully random initialization of the system dynamics and the encoder (RanDY + RanENC), 2) a BLA initialization of the system dynamics and a random initialization of the encoder (LinDY + RanENC), and 3) a BLA initialization of both the system dynamics and of the encoder (LinDY + LinENC). Table~\ref{tab:init} provides an overview of these initialization schemes.

\begin{table}[htb]
\caption{Parameter initialization scheme comparison. `Random' indicates that the values are drawn from the distribution $\mathcal{U}(-1,1)/ \sqrt{n_{in}}$ where $n_{in}$ denotes the number of function inputs.}
\begin{center}
\begin{tabular}{|p{0.8 cm}|p{1.48cm}|p{1.48cm}|p{2.8cm}|   } 
  \hline
  \textbf{} & \textbf{RanDY + RanENC}  &  \textbf{LinDY + RanENC} & \textbf{LinDY + LinENC}  \\
  \hline
  $A$ & Random  & $\tilde{A}$ & $\tilde{A}$ \\ 
  \hline
  $B$ &  Random  & $\tilde{B}$ & $\tilde{B}$ \\ 
  \hline
  $C$ &  Random  & $\tilde{C}$ & $\tilde{C}$ \\
  \hline
  $W_u$& Random & Random & $[\tilde{C}\tilde{A}^-]_\text{map}^\dagger [\tilde{C}\tilde{A}^-\tilde{B}]_\text{map}$ \\
  \hline
  $W_y$ & Random & Random & $[\tilde{C}\tilde{A}^-\tilde{B}]_\text{map}^\dagger$ \\
  \hline
  $W_\text{last}^{f,h}$ &  Random  & 0 & 0 \\ 
  \hline
  $W_\text{last}^\psi$ &  Random  & Random & 0 \\ 
  \hline
  $b_\text{last}^{f,h,\psi}$ & 0   & 0 & 0 \\ 
  \hline
\end{tabular}
\label{tab:init}
\end{center}
\end{table}

\section{Experiments} \label{ses:results}

The three initialization strategies outlined in Table~\ref{tab:init} are evaluated on a Wiener-Hammerstein (WH) simulation example, as well as on the Wiener-Hammerstein benchmark dataset \citep{schoukens2009wiener}.

\begin{figure*}[tb] 
\begin{center}
\includegraphics[width=18cm]{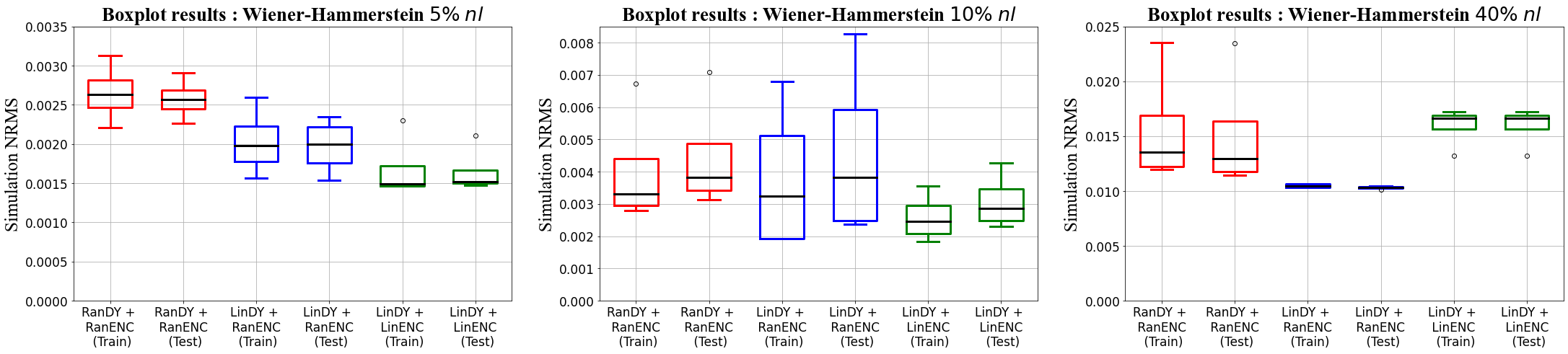} 
\caption{NRMS error of the simulated model responses computed on the training and test data sets for the WH simulation example \eqref{Equ:WH sys}. The results are displayed for the estimated models using all three initialization strategies, and 4 runs to create the box diagrams. Furthermore, the degree of non-linearity of the system is scaled from 5$\% nl$ (Left), 10$\% nl$ (Middle) to 40$\% nl$ (Right) as defined in \eqref{equ:nl-procent}.} 
\label{fig: boxplot all}
\end{center}
\end{figure*}  

\subsection{Simulation Example}
\subsubsection{System and Data:}

A SISO Wiener-Hammerstein system with a sine nonlinearity $g(x) = \sin(x)$ sandwiched between two linear low pass filters is considered. Both $G_1$, described by the state-space matrices ($A_1$,$B_1$,$C_1$), and $G_2$, represented by ($A_2$,$B_2$,$C_2$), correspond to 2nd order low-pass dynamics with a cut-off frequency at 200 Hz and 350 Hz respectively. Hence, the overall system order is 4. The state-space representation of this Wiener-Hammerstein system can be written as:
\begin{align} \label{Equ:WH sys}
    x_{t+1} &= \begin{bmatrix} A_1 & 0\\ 0 & A_2 \end{bmatrix}x_t + \begin{bmatrix} B_{1} \\ 0 \end{bmatrix} u_t + \begin{bmatrix} 0 \\ B_{2} \end{bmatrix} g(\begin{bmatrix}  C_1 & 0 \end{bmatrix}x_t)  \\
    y_t &= \begin{bmatrix} 0 & C_2 \end{bmatrix} x_t
\end{align}
To generate the data, a white Gaussian excitation for $u_t$ is considered. No noise disturbance is added to the outputs during this simulation example to emphasize the difference in model quality over the different initialization schemes. During the estimation of the BLA, the nonlinear behaviour of the system acts as a noise source \citep{pintelon2012system} and introduces variance on the estimate. The input and output signals are sampled at 1000 Hz. 150,000 data samples are obtained for the training dataset and 25,000 data samples for the validation and the test dataset.

\subsubsection{Model Structure and Hyper-Parameters:} The nonlinear terms of the encoder ($\psi_{\theta_\text{NL}}$), state ($f_{\theta_\text{NL}}$) and output functions ($h_{\theta_\text{NL}}$) are parameterized as artificial neural networks with 64 nodes, 2 hidden layers, Tanh activation functions and the non-zero elements are initialized by Xavier initialization~\citep{glorot2010Xavierinit}. The order of the model structure is set to 4. The $T$, associated with the loss function \eqref{Equ:total_loss}, is chosen to be 50 and  $n = n_a = n_b = 4$. Adam optimization with a learning rate of 0.001 and batch size of 512 is considered for both the state-space networks and encoder network. The model is trained for 500 epochs.

\subsubsection{Performance Measure:} The Normalised Root Mean Square (NRMS) of the simulation error is used as a performance measure:
\begin{equation}
    \mathrm{NRMS} = \frac{\sqrt{\frac{1}{N-n+1}\sum_{t = n}^{N}||\hat{y}_{t|n} - y_t||^{2}_{2}}}{\sigma_y},
    \label{Equ:NRMS}
\end{equation}    
where $\hat{y}_{t|n}$ is the simulated model output using the encoder to provide an estimate of the initial state and $y_t$ is the system output. $\sigma_y$ is the standard deviation of the system output in the test set.

\subsubsection{Linear model and reconstructability map:} The linear discrete time state-space model is estimated using the N4SID algorithm \citep{van1994n4sid}. The order of the linear model is set to 4, no direct feedthrough is considered. The data is preprocessed such that the input and output signals are zero-mean and have a standard deviation equal to 1.

\subsubsection{Nonlinearity Level:} The improved initialisation is tested for various levels of nonlinearity of the system. The input amplitude can be adjusted to vary the nonlinear behaviour level which will be expressed using $\%nl$ defined as
\begin{equation}\label{equ:nl-procent}
    \%nl \triangleq (1 - \mathrm{NRMS}_{BLA})\cdot 100,
\end{equation}
where the NRMS$_{BLA}$ is the NRMS simulation error~\eqref{Equ:NRMS} computed between the linear output ($\hat{y}^{BLA}$) and system output ($y$). In this experiment, there is no noise considered in the input-output dataset. Hence, the dynamics which cannot be modelled by the BLA model~\eqref{equ:BLA-ss} are due to the nonlinear system behavior. The experiments are conducted for nonlinearity percentages 1\%, 5\%, 10\%, 20\% and 40\%. 

\subsubsection{Results:} 4 independent simulation runs are performed to account for the random initialisation (see Table~\ref{tab:init}) for each of the $\%nl$ levels. 

From the simulation results in Fig.~\ref{fig: boxplot all} and Table \ref{tab:MC result}, the LinDY + LinENC initialization performs better than LinDY + RanENC and RanDY + RanENC initialization for 10$\% nl$ and below. Whereas, for 20$\% nl$ and 40$\% nl$, the LinDY + RanENC performs better than the other 2 initialisation approaches. Hence, the LinDY + LinENC initialization is especially well suited for a weakly nonlinear system.

\begin{figure}[htb]
\begin{subfigure}{\textwidth}
    \includegraphics[width= 8.8 cm]{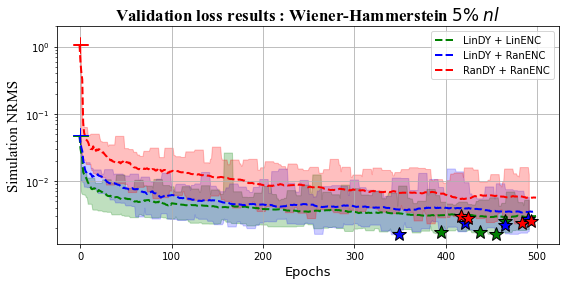}  
    \label{fig:first}
\end{subfigure}
\begin{subfigure}{\textwidth}
    \includegraphics[width=8.8 cm]{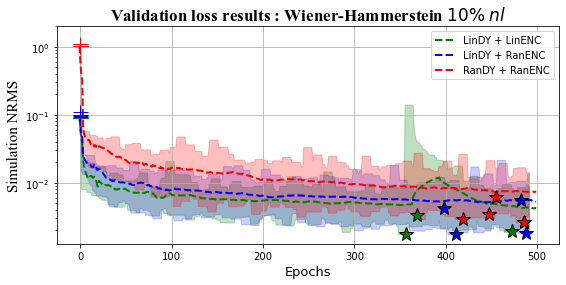}  
    \label{fig:second}
\end{subfigure}
\begin{subfigure}{\textwidth}
    \includegraphics[width=8.8 cm]{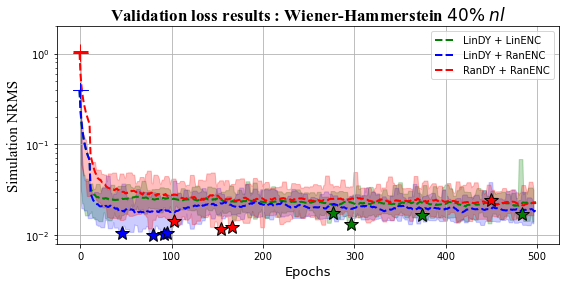}  
    \label{fig:third}
\end{subfigure}
\caption{Evolution of validation loss during the training phase for the WH simulation example. The star ($\star$) denotes the lowest validation loss for which the final model is obtained. The dashed line indicates the moving average and the shaded region denotes the variation of the validation loss over the different runs.}
\label{fig: Validation_loss}
\end{figure}

Furthermore, we can observe that using the BLA initialization of the state-space equations is always beneficial compared to a purely random initialization. This is especially observed in Figure~\ref{fig: Validation_loss}, where we observe that the LinDy + RanEnc validation loss is always lower or equal than the fully random initialization scheme.

\begin{table}
 \caption{Median of the NRMS of the simulation error of the estimated models computed on the test data set for all three initialization strategies on the WH simulation example.} 
\begin{center}
 \begin{tabular}{|p{0.7cm}|p{1.7cm}|p{1.7cm}|p{1.7cm}|}
    \hline
    \textbf{$nl\%$} & \textbf{RanDY + RanENC} & \textbf{LinDY + RanENC} & \textbf{LinDY + LinENC} \\
    \hline
    $1\%$ & 0.21\% & 0.09\% & \textbf{0.06\%} \\
    \hline
    $5\%$ & 0.26\% & 0.2\%0 & \textbf{0.15\%} \\
    \hline
    $10\%$ & 0.38\% & 0.38\% & \textbf{0.29\%} \\
    \hline
    $20\%$ & 0.94\% & \textbf{0.29\%} & 0.34\%\\
    \hline
    $40\%$ & 1.29\% & \textbf{1.03\%} & 1.63\%\\
    \hline
  \end{tabular}
\label{tab:MC result}
    \end{center}
\end{table}

Fig~\ref{fig: Validation_loss} denote the evolution of the validation loss during training. The moving average line indicates that for 10$\% nl$ and below, the LinDY + LinENC initialization provides faster convergence than the other 2 initialisation approaches. The stars ($\star$) in Fig~\ref{fig: Validation_loss} denote the best obtained model on the validation dataset. It can be noticed that in most cases, the best model is obtained towards the end of the optimization run. This suggests that better model might have obtained for more epochs. Nevertheless, the LinDY + LinENC initialization, on average, provides models of equal quality in less time for a weakly nonlinear
system $(\%nl \leq 10)$ compared to the other 2 initialisation approaches.

\subsection{Wiener-Hammerstein Benchmark}
\subsubsection{System and Data:} The Wiener-Hammerstein (WH) benchmark \citep{schoukens2009wiener} consist of a diode circuit as static nonlinearity, sandwiched between a third order Chebyshev filter and a third order inverse Chebyshev filter. The system is excited with a filtered Gaussian noise signal with a cut-off frequency of 10 kHz. In total, 80000 data-samples are used for training, 20000 for validation and 78800 for testing.

\subsubsection{Model Structure and Hyperparameters:} The encoder ($\psi_{\theta_\text{NL}}$), state ($f_{\theta_\text{NL}}$) and output functions ($h_{\theta_\text{NL}}$) are parameterised similarly to the previous experiment. The order of the model structure is set to 6. The $T$ associated with the loss function \eqref{Equ:total_loss} is set to 80 and $n = n_a = n_b = 6$. Adam optimization with a learning rate of 0.001 is considered for both the state-space networks and the encoder network. The model is trained for 3000 epochs with a batch size of 1024.

\subsubsection{Performance Measure:} The NRMS simulation error is used to assess the model performance (see~\eqref{Equ:NRMS}). 

\subsubsection{Linear Model and Reconstructability Map:} The linear model is estimated similarly to the previous experiment. The order of the estimated linear model is set to 6. The system has a nonlinearity level of about 18$\% nl$. Using the obtained 6th order LTI model, the reconstructability map \eqref{equ:map} is obtained with $n = n_a = n_b = 6$.

\subsubsection{Benchmark Results:} 
\begin{figure}[htb]
\begin{center}
\includegraphics[width=8.8 cm]{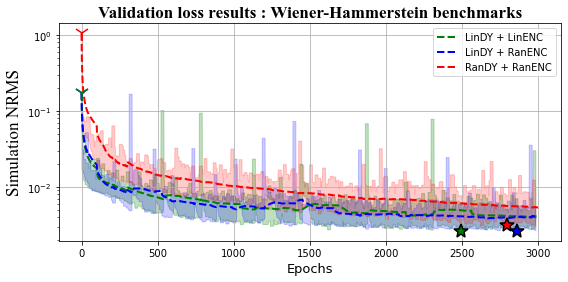}    
\caption{Validation loss during training for the Wiener-Hammerstein benchmark.} 
\label{fig:WH_bench_loss}
\end{center}
\end{figure}

It can be noticed that the lowest simulation NRMS error on the test dataset is obtained for LinDY + LinENC (see Table~\ref{tab:WH_bench}). Moreover, the LinDY + LinENC obtains its best validation result around 300 epochs before the other 2 approaches (indicated by the $\star$ in Fig \ref{fig:WH_bench_loss}), even though the moving average is very close to the LinDY + RanENC results. This is to be expected based on the previous simulation study, if we consider the 20\% nonlinearity level results obtained there. 
    
\begin{table}[htb]
    \caption{NRMS of the simulation error on the test set of the Wiener-Hammerstein benchmark.}
    \begin{center}
    \begin{tabular}{|p{3.3cm}|p{1.0cm}|}
    \hline
      \textbf{Initialisation method}  & NRMS  \\
         \hline
         RanDY + RanENC &  0.29\% \\
         \hline
         LinDY + RanENC &  0.29\% \\
         \hline
         LinDY + LinENC &  \textbf{0.25\%} \\
        \hline
    \end{tabular}
    \label{tab:WH_bench}
    \end{center}
\end{table}

\section{Conclusion}\label{sec:conclusion}
This paper has shown that the parameter initialization of the identification problem of nonlinear state-space models using the SUBNET architecture can be efficiently accomplished by the Best Linear Approximation. The state-space matrices of the linear approximate model are used as a linear bypass in the neural networks that represent the state and output equations. However, the SUBNET architecture also utilizes an encoder network that estimates the initial state of each subsection used during the network training. This encoder network acts as a reconstructability map. Hence the reconstructability map of the linear approximate model is used to initialize the encoder network. The simulation results illustrate that this is beneficial for mildly nonlinear systems.

\bibliography{ifacconf}          
\end{document}